\documentclass{aa}
\usepackage{psfig,multirow}
\thesaurus{11(11.07.1; 11.19.6); 12(12.12.1); 13(13.18.1)}
\title{The Phoenix radio survey: the angular correlation function}
\titlerunning{Correlation function of the Phoenix radio survey} 

\author{A. Georgakakis \inst{1,2}
\and B. Mobasher \inst{2}
\and L. Cram \inst{3}
\and A. Hopkins \inst{4}
\and M. Rowan-Robinson  \inst{2}}
\institute{
School of Physics and Astronomy, University of Birmingham, Edgbaston, B15
2TT, UK
\and
Astrophysics Group, Blackett Laboratory, Imperial College,
Prince Consort Rd , London SW7 2BZ, UK 
\and 
Astrophysics Department, School of Physics, University of Sydney, NSW,
Australia 2006 
\and
 Department of Physics and Astronomy, University of Pittsburgh, 3941
  O'Hara Street, PA 15260, USA}
\begin{document} 

\maketitle 
 
\begin{abstract} 

Using the Phoenix radio survey, a homogeneous survey selected at 1.4\,GHz
and covering an area of $\approx$3\,deg$^{2}$, we analyse the clustering of 
the sub-mJy radio population using angular correlation function analysis. 
Extensive simulations are carried out to investigate the significance of
the estimated angular correlation amplitudes. Our analysis show that for
the $S_{1.4}>0.5$\,mJy sub-samples the radio source distribution is
anisotropic at the 2$\sigma$ significance level. Additionally, we
estimate upper limits for the angular correlation amplitudes that, 
despite the large uncertainties, are in good agreement with 
the amplitude estimates for sources brighter than 1\,mJy, detected
in the FIRST radio survey (Cress {et al.} 1997).  Adopting a radio
luminosity function and assuming an evolving spatial correlation
function of the form  $\xi(r)=(r/r_{o})^{-\gamma}
(1+z)^{-(3+\epsilon)}$, with the evolution parametrised by $\epsilon$,
we  find an upper limit for the angular correlation length
$r_{o}\approx9$\,h$^{-1}$\,Mpc  for $S_{1.4}>0.5$\,mJy and
$\gamma$=2.1. This agrees well with the value
$r_{o}\approx6-8$\,h$^{-1}$\,Mpc estimated from the FIRST  radio
survey for sources brighter than 1\,mJy. Additionally, we  quantify
the characteristics, in terms of areal coverage and limiting flux
density, of future deep radio surveys to yield a significant
correlation amplitude detection and to explore possible changes of the
correlation amplitude with flux density. 

\keywords{
Galaxies: general -- Galaxies: structure -- Cosmology: large scale
structure of Universe -- Radio continuum: galaxies}
 
\end{abstract} 
 
\section{Introduction}\label{introduction}
The spatial and angular correlation functions, $\xi(r)$ and  $w(\theta)$
respectively,  have been extensively used to quantify the clustering of
galaxies. The former has the important property of being the Fourier
conjugate of the power spectrum, $P(k)$, of the galaxy distribution
(e.g. Peacock \& Nicholson 1991). The latter is the two dimensional
projection of $\xi(r)$ on the sky and is related  to it by the Limber
equation  (Limber 1953; Phillipps {et al.} 1978). Although $w(\theta)$
conveys less information on galaxy clustering than $\xi(r)$ due to its
integral nature, nevertheless  it is a powerful probe of the large scale
distribution of galaxies, especially at faint magnitudes where redshift
information is difficult to obtain.

Wide-area surveys at optical wavelengths have provided information on the 
large-scale structure at redshifts $z\approx0.1$ (Loveday {et al.}
1995; Maddox {et al.} 1990b). Additionally, deep optical and
near-infrared  samples over smaller areas (Villumsen {et al.} 1997;
Hudon {et al.} 1996; Roche {et al.} 1998a, b; Carlberg {et
al.} 1997) give useful information on the large-scale distribution of
galaxies at higher redshifts ($z\le$1.0). There is a general consensus 
that at optical wavelengths, $w(\theta)$ is a power law of the form
$w(\theta)=A_{w}\,\theta^{-(\gamma-1)}$, with $\gamma\approx1.8$ (Maddox
{et al.} 1990b). The amplitude, $A_{w}$, decreases with increasing depth
of the survey, while there is also evidence that the slope, $\gamma$,
flattens at faint magnitudes (Infante \& Pritchet 1995). The spatial
correlation function has the  form $\xi(r)=(r/r_{o})^{-\gamma}$,  where the
correlation length, $r_{o}$, is found to be $r_{o}=5.4$\,h$^{-1}$\,Mpc
(Davis \& Peebles  1983). However, it has been demonstrated that
early-type galaxies are more clustered (higher values of $r_{o}$) than late
types and that lower luminosity galaxies are a factor of $\approx2$ less
clustered than their brighter counterparts of the same Hubble type (Loveday
{et al.} 1995). There is also evidence that $r_{o}$ decreases with
increasing depth of the survey  (Infante \& Pritchet 1995), due to the
presence of a population of weakly clustered faint galaxies.   

Radio surveys, unlike the optical ones, are not affected by galactic dust
extinction and mainly consist of high redshift objects
($\overline{z}\approx1$). Therefore, they  sample much larger volumes  than
optically selected samples, albeit with sparser coverage, providing the
opportunity  to study the clustering of matter at much larger physical
scales. However, the broad luminosity function of extragalactic radio
sources (Condon 1984; Dunlop \& Peacock 1990)  implies that a
flux density-limited sample spans a wide range of redshifts. The projection
on the sky  of all detected objects results in a distribution close to
random, smearing out any information on the large-scale structure. This is
supported by the fact that no  signal has been detected  in the angular
(2-D) correlation analysis of bright radio galaxies (Webster 1976; Masson
1979). However, Shaver \& Pierre (1989) found evidence of anisotropic
distribution  of radio sources towards the   supergalactic plane extending
out to at least $z\approx$0.02. More recently, a clustering length of
$r_{o}=11$\,h$^{-1}$\,Mpc was estimated at 1.4\,GHz from the spatial (3-D)
correlation analysis of radio sources having flux densities $>0.5$\,Jy and
redshifts in the  range $0.01<z<0.1$ (Peacock \& Nicholson 1991).  
 
Cress {et al.} (1996) have suggested that at  lower flux densities,
projection effects may become less significant and hence, a 2-D correlation
analysis can be applied successfully to investigate  the clustering of
radio galaxies. Indeed, a non-zero amplitude has been found for the angular
correlation function of radio sources with  $S_{4.85GHz}>50$\,mJy in both
the Green Bank (northern hemisphere) and the Parkes-MIT-NRAO (southern
hemisphere) 4.85\,GHz sky surveys  (Loan, Wall \& Lahav 1997; Kooiman,
Burns \& Klypin 1995). More recently, using the FIRST radio survey (1.4\,GHz;
Becker, White \& Helfand 1995), with a uniform sensitivity of 1\,mJy over
an area of 1500\,deg$^{2}$, a non-zero and clearly significant  amplitude
for the angular correlation function is estimated (Cress {et al.}
1996). Adopting the radio luminosity functions (RLF) determined
independently by Condon (1984) and Dunlop \& Peacock (1990), Cress {et
al.} (1997) inferred a clustering  length of $r_{o}$=6-8\,h$^{-1}$\,Mpc for
$S_{1.4}>$1\,mJy radio sources.   

The recent large-area  radio surveys described above, have provided
information on the two-dimensional projected distribution of relatively
bright ($S_{1.4}\ge$1\,mJy) radio sources, dominated by
bright ellipticals and AGNs. However, there is still limited
information on the clustering properties of the faint (sub-mJy)  
radio population. 
At flux densities below few mJy there is evidence for the 
appearance of a new population of faint radio sources, likely to comprise a
large fraction of starbursts (Benn {et al.} 1993; Georgakakis {et al.} 
1999). In particular, the radio luminosity function models  developed by
Condon (1984) predict that the surface  density of starbursts increases
from $10\%$ of the radio population at  $S_{1.4}\approx1$\,mJy to over
$30\%$ at $S_{1.4}\approx0.4$\,mJy.
Moreover, studies of the spatial distribution of starbursts, selected at
optical and infrared wavelengths, show that these objects are expected to
have different clustering properties (Davis \& Geller 1976; Giovanelli {et
al.} 1986; Saunders, Rowan-Robinson \& Lawrence 1992; Loveday {et al.}
1995) compared to those of early-type galaxies, with which bright radio
sources are often associated. Therefore, study of the angular correlation
function at sub-mJy flux density levels has the potential to reveal
differences in the clustering properties of the faint radio
population. 

Recently, Benn \& Wall (1995) demonstrated that the isotropy (or
anisotropy) in the radio source counts in different fields of similar
geometry and sensitivity can be used to set limits on the scale of the
largest cellular structures in the Universe. However, they argued that at
sub-mJy and $\mu$Jy flux densities, such a study is hampered by the small
number of comparable surveys and by the small number of sources detected in
each field. This is also supported by Windhorst {et al.} (1990), who
reviewed the field-to-field variations in the radio source counts of small
area radio surveys ($\le1\mathrm{\,deg^{2}}$). They concluded that the
differences, although exceeding the random distribution expectation, are
due to statistical fluctuations, rather than of cosmic
nature. Nevertheless, a non-uniform angular correlation function was
estimated by Oort (1987a) for the radio sources detected in the deep
($S_{1.4}>0.1$\,mJy) Lynx fields (Oort 1987b). More recently, Richards et
al. (1999) also reported the detection of clustering signal for radio
sources brighter than $40\,\mu$Jy, detected within a $40$\,arcmin
diameter radio survey (1.4\,GHz) centred on the HDF.

The Phoenix radio (1.4\,GHz) survey, covering an area of 3.14\,deg$^{2}$,
larger than any other survey at a similar flux density limit
($S_{1.4}=0.4$\,mJy) and reaching surface densities $\approx 6.6\times 
10^{5}\mathrm{\,sources\,sr^{-1}}$, provides a unique opportunity to  study
the clustering properties of the faint radio population. In this paper the
angular correlation function of the radio sources detected in the Phoenix
field is estimated. Section 2 gives a brief description of the
observations. The method for calculating $w(\theta)$  is outlined in
section 3. In section 4 we estimate the  correlation function of the radio
sample. Section 5 presents the simulations carried out to investigate the
significance of our results, while in section 6 the correlation function
amplitudes are  compared with 3-D clustering models. The results from the
radio correlation analysis are discussed in section 7. Finally, we
summarise our conclusions in section 8. 
 
\section{Radio Observations}\label{observations}

The Phoenix survey has been performed at 1.4\,GHz using the 6A
configuration  of the Australia Telescope Compact Array (ATCA). It is a
mosaic of 30  pointing centres covering a $2^{\circ}$ diameter field
centred at   $\mathrm{RA}(2000)=01^{\mathrm h}~14^{\mathrm m}~12\fs16$ ;
$\mathrm{Dec.}(2000)=-45^{\circ}~44'~8\farcs0$. The final synthesised beam
size is $\approx$8\,arcsec. 
  
Details of the observation, data reduction, source detection and 
catalogue generation are presented in Hopkins {et al.} (1998).  
A source is included in the sample if its peak flux density 
is 4$\sigma$ above the local RMS and if it survives visual inspection.  
There are two sources of incompleteness in the catalogue, 
as with any peak flux density limited sample. The first is a result of the
attenuation of the  primary beam sensitivity away from a field centre. This
effect has  been minimised in the Phoenix survey by the mosaicing strategy
used. The second is the fact that extended objects with a total flux
density above the survey limit can be missed by a source detection
algorithm which initially detects candidates based on their peak flux  
densities. These effects have been described in Hopkins {et al.}
(1998). The catalogue is estimated to be $50\%$ complete to 0.3\,mJy,
and $80\%$ complete to 0.4\,mJy. Therefore  to minimise the effect  of
incompleteness when performing the correlation  analysis, we restrict
ourselves to a subsample  containing all sources brighter than 0.4\,mJy.  
 
\section{Two point correlation function}\label{sec_3}

The two-point angular correlation function, $w(\theta)$, is defined as the
joint  probability $\delta P$ of finding sources within the solid angle
elements $\delta \Omega_{1}$ and  $\delta \Omega_{2}$, separated by an
angle~$\theta$, in the form
 
\begin{equation}\label{eq_1}   
 \delta P= N^{2}\, (1+w(\theta)) \delta \Omega_{1} \delta \Omega_{2},
\end{equation} 

\noindent where $N$ is the mean surface density of galaxies. For a random
distribution of sources $w(\theta)$=0. Therefore, the  angular correlation
function provides a measure of galaxy density excess over that expected for
a random distribution. Various methods for estimating $w(\theta)$ have been
introduced, as  discussed by Infante (1994). In the present study, a source
is taken as the `centre' and the number of pairs within annular rings is 
counted. To account for the edge effects, Monte Carlo techniques are used by
placing random points within the area of the survey. The most commonly used
estimators in this procedure have the form 

\begin{equation}\label{eq_2}  
w=\frac{DD}{RR} - 1,\; {\mathrm or} 
\end{equation} 
\begin{equation}\label{eq_3}  
w=\frac{DD}{DR} - 1,
\end{equation} 
 
\noindent where DD and DR are respectively the number of sources and random
points computed  at separations $\theta$ and  $\theta + d\theta$ from a
given galaxy. Similarly, RR is the number of random points within the
angular interval $\theta$ to $\theta + d\theta$ from a given random point.
Infante (1994) and Hewett (1982) emphasised the importance of correcting
for any spurious cross correlation between the random and the galaxy
catalogues when using the above estimators, arising partly because of the
anisotropic and inhomogeneous galaxy distribution relative to the field
boundaries. The corrected form of the estimator is then  
 
\begin{equation}\label{eq_4}  
w=\frac{DD}{DR} - \frac{RD}{RR},
\end{equation}  

\noindent where RD is the number of random-data pairs, taking the random
points as centres (Infante 1994). Landy \& Szalay (1993, hereafter LS)
introduced the estimator  
 
\begin{equation}\label{eq_5}  
w=\frac{DD-2DR+RR}{RR}.
\end{equation}  
 
\noindent The advantages of this estimator is that it minimises the
variance  of $w(\theta)$ and also reduces the edge effects from which
both the estimators (\ref{eq_3}) and (\ref{eq_4}) suffer. Moreover, it
has similar properties (Landy \& Szalay 1993) to that introduced by
Hamilton (1993)    
 
\begin{equation}\label{eq_6}  
w=\frac{DD\, RR}{DR\, DR}-1.
\end{equation}
 
\noindent In the present study the correlation function has been calculated
using estimators (\ref{eq_4}), (\ref{eq_5}) and (\ref{eq_6}). The
results agree within the respective uncertainties. Therefore, only 
the  results from the LS estimator are presented.  
 
For a given radio galaxy sample, a total of 100 random catalogues were
generated, each having the same number of points as the original data
set. The random sets were cross-correlated with the galaxy catalogue,
giving  an average value for $w(\theta)$ at each angular separation. In
producing random sets of points, we take into account variations in
sensitivity, which might affect the correlation function estimate. The flux
density threshold for detection depends on the local RMS noise which varies
across the survey area. Since the random fields are expected to have the  
same sources of bias as the data (i.e. the simulated catalogues must have
the same selection effects as the real catalogue), the RMS noise map, with
a resolution of $\approx$10\,arcsec, is used to discard simulated points in
noisy areas. This is accomplished, to the first approximation, by assigning
a flux density to random points using the Windhorst {et al.} (1985)
log$N$-log$S$ distribution, following the method of Cress {et al.}
(1996). If the flux density of the random point is less than 4 times the
local RMS noise, the point is excluded from the random data set. However,
producing random points with a uniform distribution over the observed area
does not change our final result.  

The uncertainty in $w$($\theta$) is determined using both Poisson
statistics and 50 bootstrap resamples of the data (Ling, Frenk and Barrow
1986). For the latter method, simulated data-sets were generated by
sampling N points with replacement from the true data-set of N points. The
correlation function is then calculated for each of the bootstrap samples,
following the same procedure as that with the original data-set. The
standard deviation  around the mean is then used to estimate the
uncertainty in the correlation function.  

Although the LS estimator used here is shown to have Poissonian
variance for uncorrelated points (Landy \& Szalay 1993), it does not
necessarily behave this way for correlated data. The bootstrap method
is believed to give a more representative estimate of the  uncertainty
associated with $w(\theta)$. However, Fisher et al. (1994) carried out
a detailed study of the biases affecting the bootstrap errors
(e.g. cosmic variance, sparse sampling by galaxies of the
underlying density distribution) and concluded that overall the
bootstrap uncertainties overestimate the true errors. 
Nevertheless, the bootstrap resampling is a general method for
assessing the accuracy of the angular correlation function estimator and
it will be used here to calculate the formal errors. The uncertainty
associated with $w(\theta)$, estimated by the bootstrap resampling
technique, is found to be about three times larger than the Poisson
estimate.

\begin{table} 
\footnotesize 
\begin{center} 
\begin{tabular}{cc} 
Flux density (mJy) & Number of Sources \\  
 $>$0.4             &  634   \\  
 $>$0.5             &  529   \\  
 $>$0.6             &  454   \\  
 $>$0.7             &  391   \\  
 $>$0.9             &  316   \\ 
 $[$0.4,0.9$]$      &  318   \\ 
\end{tabular} 
\end{center} 
\caption{Number of sources in each flux density-limited 
subsample}\label{tab_1}
\normalsize  
\end{table} 

There is expected be a significant number of physical double sources in a
radio survey with the angular resolution of the Phoenix survey
($\approx8^{\prime\prime}$). In any study of the clustering of radio
galaxies via correlation  analysis, these should not be considered as two
sources since both components are formed in the same galaxy. To identify
groups of sources that are likely to be sub-components of a single source
we have employed a percolation technique where all sources within a given
radius are replaced by a single source at an appropriate `centroid' (Cress
{et al.} 1996;  Magliocchetti {et al.} 1998). Following the method
developed by Magliocchetti {et al.} (1998), we vary the link-length in the
percolation procedure according to the flux of each source. In that way,
bright sources are combined, even if their angular separation is large,
whereas faint sources are left as single objects. This technique is based
on the $\theta-S$ relation for radio sources, which has been shown to
follow $\theta \propto \sqrt S$ (Oort 1987a). 

\begin{figure} 
\centerline{\psfig{figure=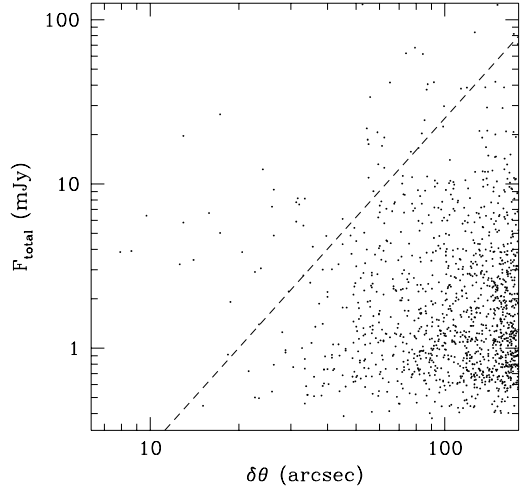,width=0.45\textwidth,angle=0}} 
\caption{Angular separation against total flux density 
of double sources. The dashed line represents the maximum link-length, 
for a given flux density, used
in the percolation technique.}\label{fig_1}
\end{figure}

To define the relation between flux density and link-length, the angular
separation of double sources versus their total flux is plotted in Figure
\ref{fig_1}, out to a separation of 180\,arcsec. Visual inspection has
confirmed that the pairs on the left of  Figure \ref{fig_1} are
predominantly sub-components of a single source. This is based  on an
assessment of the appearance of the object, including the disposition of
the sources, the nature of any bridging radiation, and the appearance of
source edges. Accordingly, we set the maximum link-length to be 

\begin{equation}\label{eq_7}  
\theta_{link} = 20 \times \sqrt{F_{total}},
\end{equation} 

\noindent where  $F_{total}$ is the total flux of each group. This is shown
by the dashed line in Figure \ref{fig_1} and effectively removes the
majority of visually identified doubles. Moreover, one can apply an
additional criterion to identify genuine doubles, based on the relative
flux densities of the sub-components (Magliocchetti {et al.} 1998). This is
because lobes of a single radio source are expected to have correlated flux
densities. Here, the groups of sources identified by the percolation
technique are combined only if their fluxes differ by a factor of less than
4. This procedure was repeated until no new groups were found. The final
catalogue consists of 908 objects to the limit of 0.1\,mJy, with a total of
30 groups of sources being identified and replaced.  

\begin{figure} 
\centerline{\psfig{figure=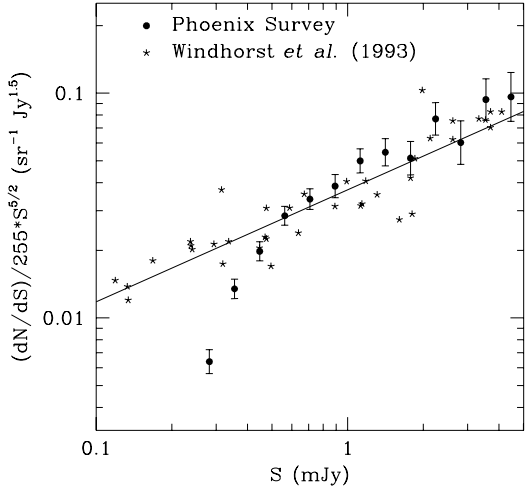,width=0.45\textwidth,angle=0}} 
\caption{Circles: source counts at 1.4\,GHz for the Phoenix Survey;
Stars: source counts from Windhorst {et al.} (1993); Line:
best fit to the Windhorst {et al.} (1993) source counts 
for flux densities less than 5\,mJy.}\label{fig_2}
\end{figure}

The source counts of the sample, normalised to the Euclidean slope, are
plotted in Figure \ref{fig_2} along with the radio counts at 1.4\,GHz
taken from  Windhorst {et al.} (1993). There is a drop in the number
counts  below 0.5\,mJy, as our sample is affected by incompleteness. To
quantify this, we firstly fit a straight line to the source counts of
Windhorst {et al.} (1993) for flux densities fainter that 5\,mJy
(continuous line in Figure \ref{fig_2}). We then compare our number
counts at a given flux density bin with those predicted by the fitted
line. We conclude that our sample is $\approx80\%$  complete to
0.4\,mJy. This is in agreement with the correction factors for
incompleteness independently derived  by Hopkins {et al.} (1998) accounting
for both resolution effects and the attenuation of the beam away from the
field centre. Therefore, to minimise the effect of incompleteness of the
radio catalogue, when performing the correlation analysis, we restrict
ourselves to a subsample  containing all sources brighter than 0.4\,mJy.

Finally, before fitting a power law to $w(\theta)$, we take into account a
bias arising from the finite boundary  of the sample. Since the angular
correlation function is calculated within a  region of solid angle
$\Omega$,  the background projected density of sources, at a given flux
density limit $S_{o}$, is effectively $N_{s}(S>S_{o})/\Omega$ (where
$N_{s}(S>S_{o})$ is the number of detected objects  brighter than $S_{o}$).
However, this is an overestimation of the true underlying mean surface
density, because of the positive correlation between  galaxies in small
separations, balanced by negative values of $w(\theta)$ at larger
separations. This bias, known  as the integral constraint, has the effect
of reducing the amplitude  of the correlation function by

\begin{equation}\label{eq_8}
\omega_{\Omega}=\frac{1}{\Omega ^{2}} \int{\int{w(\theta) d\Omega_{1} d\Omega_{2}}},  
\end{equation} 
 
\noindent where $\Omega$ is the solid angle of the survey area. The angular
constraint is estimated using Monte Carlo integration, assuming   the
correlation function to follow a power law $w(\theta)=A_{\omega} \times
\theta ^{-\delta}$  with $\delta$=0.8 or 1.1. We find
$\omega_{\Omega}$=1.46$A_{\omega}$ and $\omega_{\Omega}$=1.89$A_{\omega}$
for $\delta$=0.8 and $\delta$=1.1 respectively, where $\theta$ is measured
in degrees. 

\section{Correlation Function of the Phoenix radio survey}\label{sec_4}

\begin{figure*} 
 \centerline{\psfig{figure=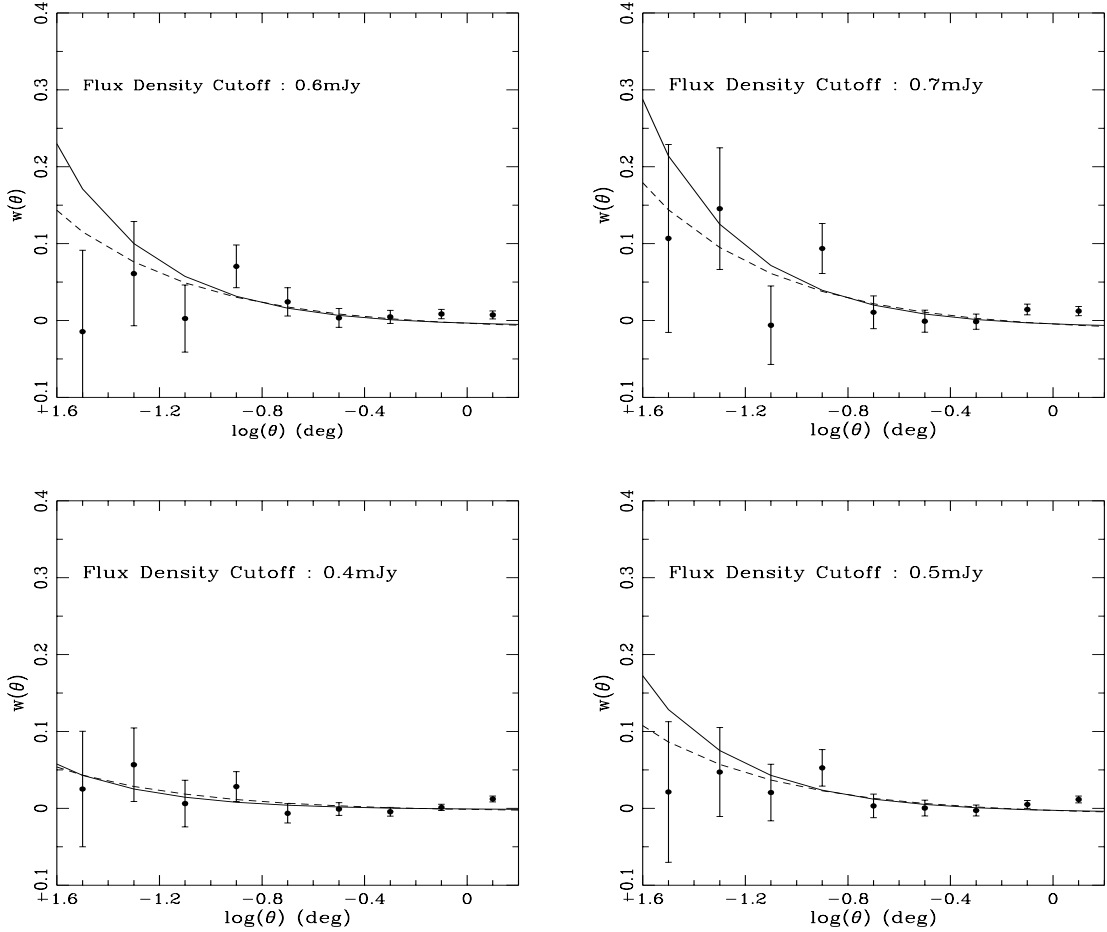,width=6in,height=5in,angle=0}} 
 \caption {Angular correlation function $w(\theta)$ of the Phoenix radio
 survey for subsamples with flux density cutoffs at 0.4, 0.5, 0.6 and
 0.7\,mJy. The error bars shown are Poisson estimates. The solid and dashed
 lines represent the fits to the data assuming $\delta$=0.8 and $\delta$=1.1
 respectively. Points in the range  $3^{\prime} <\theta <20^{\prime}$ are
 used  in the fitting algorithm.}\label{fig_3} 
\end{figure*}

\begin{figure*} 
\centerline{\psfig{figure=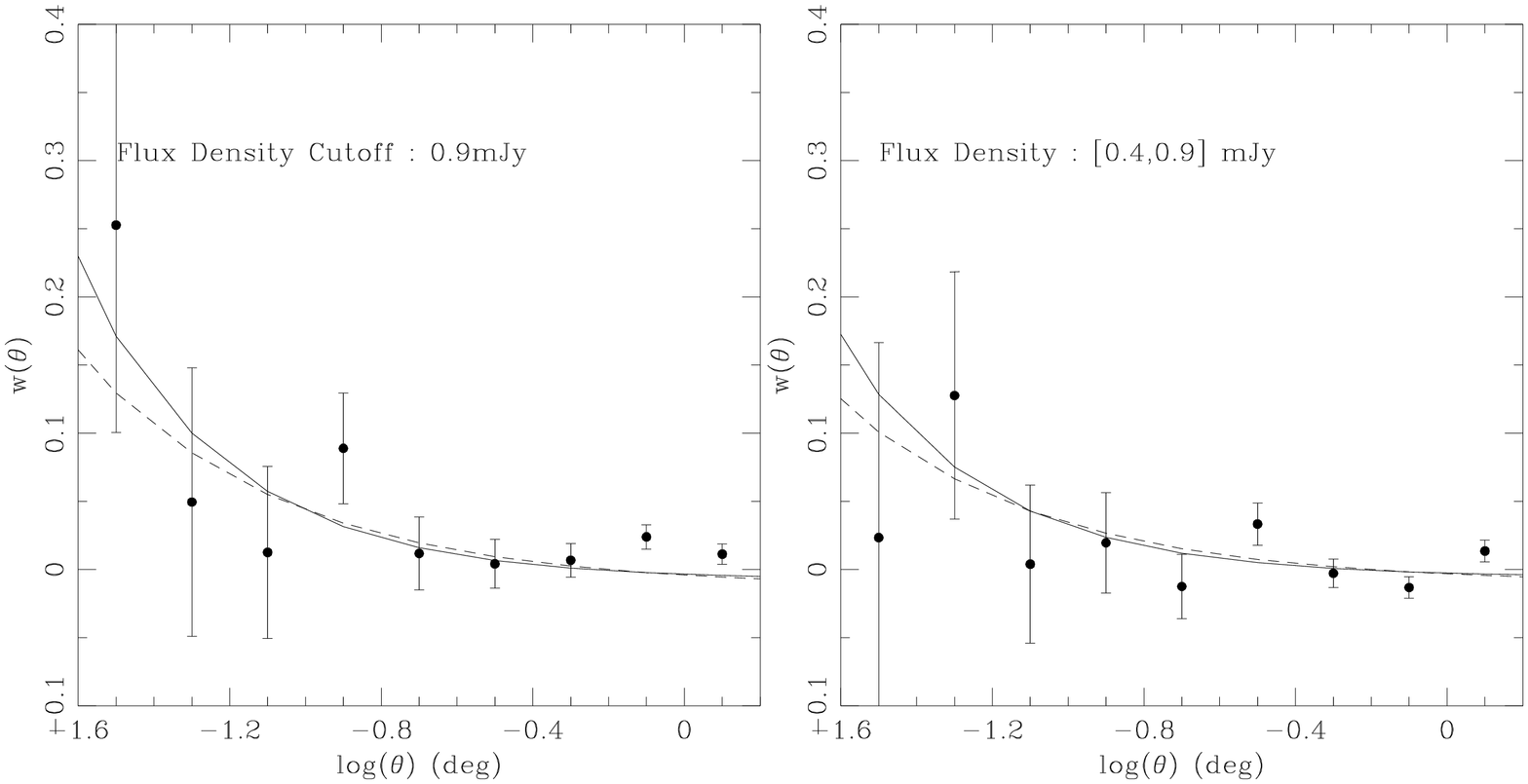,width=6in,height=3in,angle=270}} 
 \caption {Angular correlation function for sources with flux densities 
 $S_{1.4}>$0.9\,mJy and $0.4<S_{1.4}<0.9$\,mJy. The error bars shown are Poisson
 estimates. The solid and dashed lines represent the fit to the data
 assuming $\delta$=0.8 and $\delta$=1.1 respectively. Points in the range
 $3^{\prime} <\theta <20^{\prime}$ are used in the fitting
 algorithm.}\label{fig_4}  
\end{figure*}

We have extracted from the original catalogue four flux limited samples
with flux density cutoffs at 0.4, 0.5, 0.6 and 0.7\,mJy. Furthermore, to 
investigate changes in the clustering  properties of radio sources with
flux density, two independent sub-samples were considered, having the same
number of objects and flux densities in the range $0.4<S_{1.4}<0.9$\,mJy and
$S_{1.4}>$0.9\,mJy. Table \ref{tab_1} lists the number of sources in each
sub-sample. The correlation function $w(\theta)$ is calculated  for each
sub-sample for angular separations  ranging from $0.03$\,deg to $1.3$\,deg,
within 9 equally separated logarithmic bins.  The results are shown in
Figures \ref{fig_3} and \ref{fig_4}.

The amplitude of the correlation function is estimated by fitting the
following function to the observations  

\begin{equation}\label{eq_9} 
 w(\theta)=A_{w} \times \theta^{-\delta}-\omega_{\Omega}(\delta). 
\end{equation} 

\noindent where the integral constraint is evaluated for the two different
values of $\delta$ adopted in previous studies; $\delta$=0.8 (Peacock \&
Nicholson 1991; Loan, Wall \& Lahav 1997) and $\delta$=1.1 (Cress {et
al.} 1996). The calculated amplitudes and the Poissonian uncertainties  
are listed in  Table \ref{tab_2}. Assuming Poisson errors, these
amplitudes are non-zero  at the 2$\sigma$ confidence level for all
the sub-samples, except from those with $S_{1.4}>0.4$\,mJy and
$0.4<S_{1.4}<0.9$\,mJy. However, since the formal bootstrap errors are
$\approx3$ times larger than the Poisson expectation, the amplitudes
derived here are treated as upper limits.  

Additionally, there are two effects that might further increase the
correlation amplitude errors: (i) the inter-dependence of the 
$w(\theta)$ measurements at different angles  $\theta$ and (ii) the
dependence of the uncertainty estimates on higher order
correlations (skewness and kurtosis) that are ignored in this study. 
However, Mo et al. (1992) demonstrated that the underestimation  of the
correlation amplitude uncertainty, $\delta A_{w}$,  by these two effects is
compensated by the overestimation of the $w(\theta)$ errors by the
bootstrap resampling method. 

Furthermore, because of the small solid angle of the Phoenix field, cosmic
variance may affect the estimated correlation  amplitudes. The effect of
cosmic variance is studied in detail in the next section.

\begin{table} 
 \footnotesize 
 \begin{center} 
 \begin{tabular}{ccc}  

 Flux Density (mJy) & $A_{w}^{(\delta=1.1)}$ & $A_{w}^{(\delta=0.8)}$ \\ 
 $>$0.4       & 0.001$\pm$0.001       &  0.003$\pm$0.003     \\  
 $>$0.5       & 0.003$\pm$0.001       &  0.006$\pm$0.003     \\  
 $>$0.6       & 0.004$\pm$0.002       &  0.008$\pm$0.004     \\  
 $>$0.7       & 0.005$\pm$0.002       &  0.010$\pm$0.004     \\  
 $>$0.9       & 0.004$\pm$0.002       &  0.009$\pm$0.005     \\ 
 $[$0.4,0.9$]$ & 0.003$\pm$0.002       &  0.007$\pm$0.005     \\ 
 \end{tabular} 
 \end{center} 
 \caption{Angular correlation function amplitudes and the associated Poisson
  errors.}\label{tab_2}
 \normalsize  
\end{table}

\section{Simulations}\label{sec_5}

Because of the relatively small Phoenix field size, one should be cautious
about the interpretation of the correlation amplitude upper limits,
since field-to-field fluctuations may affect our results. For example
Oort (1987a) carried out the angular correlation function analysis
using sub-mJy radio surveys smaller than Phoenix and found evidence
for anisotropic source distribution in only some of them.  

\begin{figure} 
 \centerline{\psfig{figure=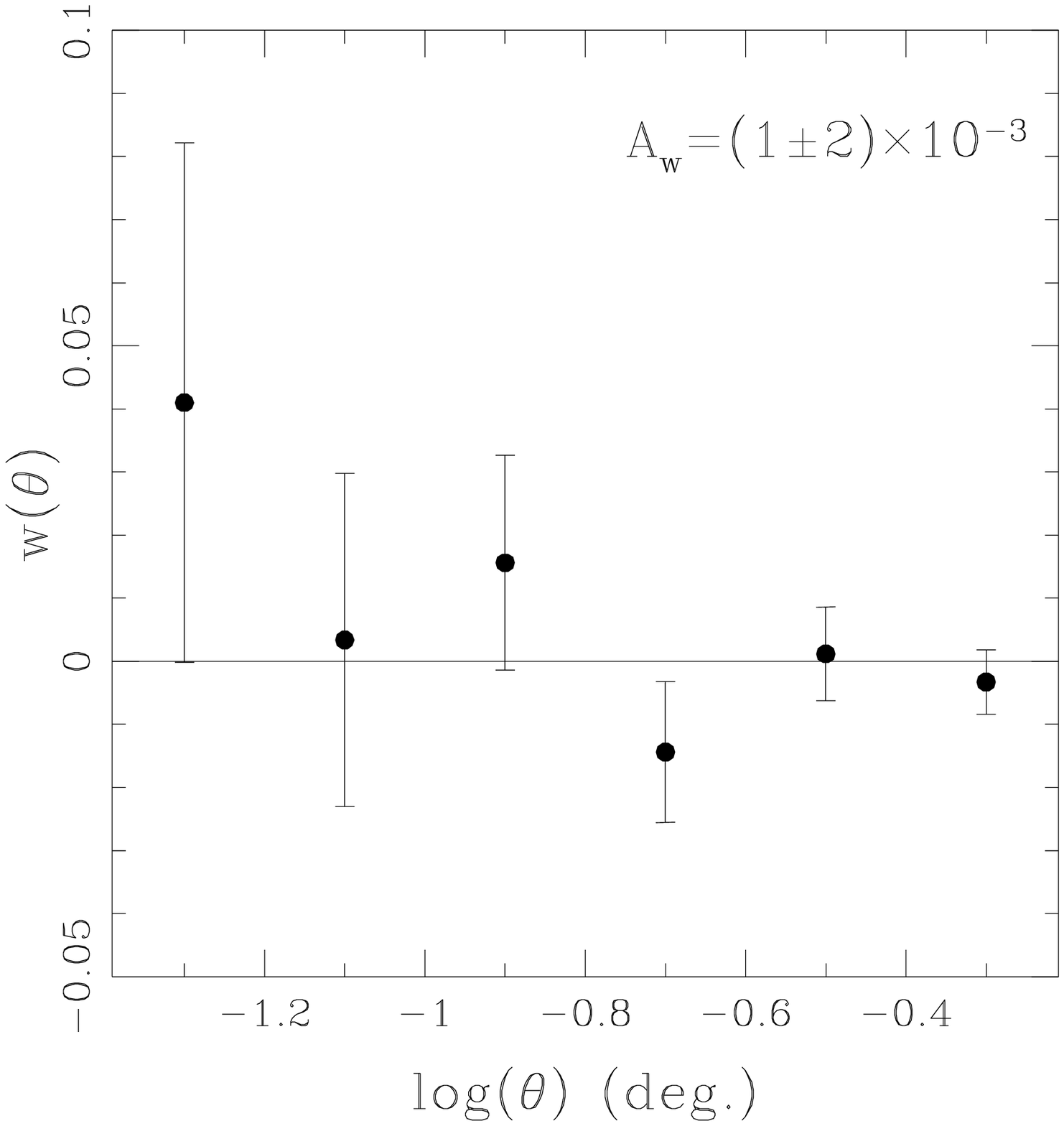,width=0.45\textwidth,angle=0}}  
 \caption[Simulations: $w(\theta)$ for a sample of random points]
 {The angular correlation function for a random sample of
 points. The error bars are Poisson estimates.}\label{fig_5} 
\end{figure}

Firstly, to test the significance of our results we construct catalogues of
randomly distributed points and calculate the correlation function from a
region with the same geometry as the Phoenix field. In all trials, the
estimated amplitudes are found to be consistent with zero within the 
Poisson standard deviation, $\sigma_{Poisson}$ (Figure \ref{fig_5}). 
Therefore, the amplitudes derived here  (non-zero at the
2\,$\sigma_{Poisson}$ significance level), imply a non-uniform distribution
of the faint radio population, albeit at the $2\sigma$ confidence
level. The exceptions are the  $S_{1.4}\ge 0.4$\,mJy and $0.4\le S_{1.4} \le
0.9$\,mJy sub-samples. This might indicate a low correlation amplitude at
faint flux densities  (e.g. $A_{w}\le 3\times10^{-3}$, $\gamma=1.8$,
$S_{1.4}\ge 0.4$\,mJy), that cannot be detected by the small size of the
Phoenix field. However, one should be cautious about this interpretation,
since at 0.4\,mJy the sample  is also likely to be affected by
incompleteness.  

To further investigate the sensitivity of our results to cosmic variance,
artificial galaxy catalogues are constructed with a built-in correlation
function of the  form $A_{w}\,\theta^{-\delta}$. Then, we attempt to
recover the correlation amplitude from a field having the same geometry as
the Phoenix survey.

The algorithm used to generate  simulated catalogues with projected
hierarchical power-law clustering is described by Infante (1994) and is
similar to that employed by Soneira \& Peebles (1978). Firstly, two points
separated by an angle $\theta_{1}$ are randomly placed within a
50$\times$50 degrees area. Each of these points serves as the center of a
new pair of randomly oriented points with angular separations
$\theta_{2}=\theta_{1}/\lambda$, where $\lambda$ is a real number. The four
points generated at the previous step are the new centers for pairs of
points separated by $\theta_{3}=\theta_{2}/\lambda$. Therefore the $L$th
step produces $2^{L-1}$ pairs with angular distance $\theta_{L}$ between
the points of the pair.  

\begin{figure*}
\centerline{\psfig{figure=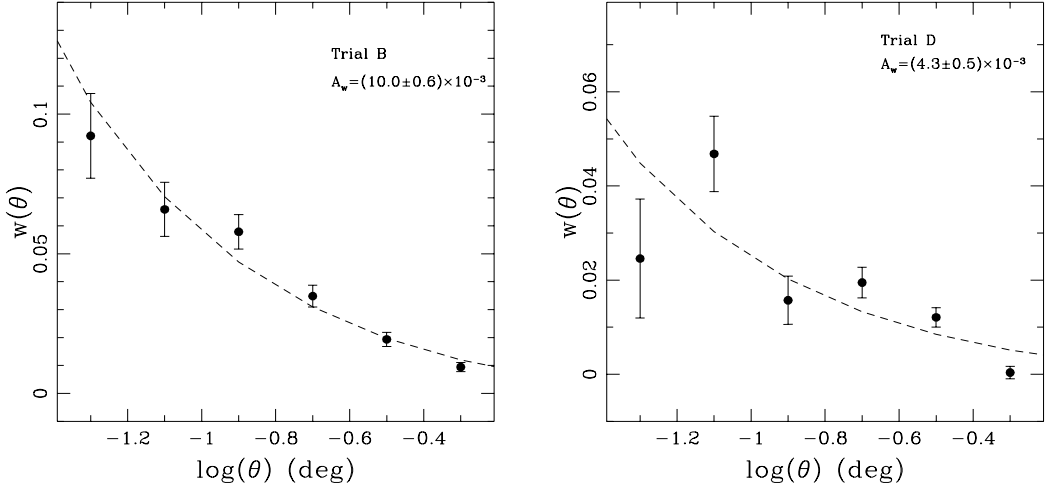,width=6in,height=3in,angle=270}} 
 \caption[Simulations: $w(\theta)$  for trials $C$ and $D$]
 {The angular correlation function for the simulations $B$ and $D$
 calculated from an area $\approx10$ times larger than that of the Phoenix  
 field. The error bars are Poisson estimates.}\label{fig_6} 
\end{figure*}

Adopting a value $\lambda=\theta_{l}/\theta_{l+1}=1.8$ for the ratio of
separations between successive levels, produces a power-law
distribution of points with exponent $\delta\approx0.8$ (Soneira \& Peebles
1978). The angular distance between points at the first level is set
$\theta_{1}$=1.5\,deg. Additionally, the number of hierarchical levels,
$L$, is drawn from a uniform probability distribution in the interval
$[L_{min},L_{max}]$=[6,9]. The minimum and maximum number of levels are
such that the smallest angular separation of the points of a pair is less
than $\approx$1\,arcmin. However, these parameters produce too many points
at each level, resulting  in large angular correlation amplitudes, $A_{w}$,
compared to  those found in bright radio samples  (Loan, Wall \& Lahav 1997;
 Cress {et al.} 1996).  Therefore, each point is assigned a fractional
survival probability, $f$. This step does not change the form of the
built-in angular correlation function (Bernstein 1994), but allows tuning
of $A_{w}$. Values of $f$ between  0.01 and 0.05 produce amplitudes in the
range $1\times10^{-3}$ to  $20\times10^{-3}$ respectively. Here we only
consider amplitudes in the interval $1\times10^{-3}<A_{w}<10\times10^{-3}$,
similar to those found in brighter radio samples (Cress {et al.} 1996). The
parameters used for four such simulations (trials $A$-$D$) are listed in 
Table \ref{tab_3}. The estimated $w(\theta)$ for trials $B$, $D$ are
plotted in Figure \ref{fig_6}.

\begin{figure} 
\centerline{\psfig{figure=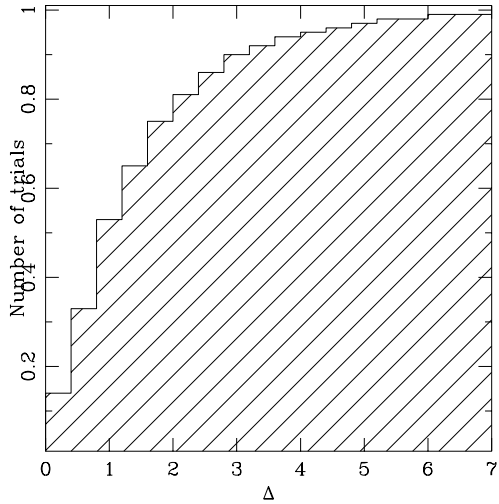,width=0.45\textwidth,angle=0}} 
 \caption[Cumulative distribution of $\Delta$]
 {Cumulative distribution of $\Delta$ defined in equation
 \ref{eq_10}.}\label{fig_7} 
\end{figure}

The procedure described above is repeated until the surface density  of points
within the 50$\times$50 degrees area exceeds the surface density of sources
of the present radio sample, at a given flux density cutoff. The angular
correlation function is then calculated from the central 36\,deg$^{2}$
region, to ensure a uniform surface  density of points. The amplitude of
the correlation function is estimated  by fitting the following function to
the observations

\begin{table} 
\footnotesize 
 \begin{center} 
 \begin{tabular}{cccccc}  

 Trial  & $\lambda$ & $L_{min}$ & $L_{max}$ & $f$  & density (deg$^{-2}$) \\
  $A$   & 1.8       &  6       &   9       & 0.03 &   145               \\
  $B$   & 1.8       &  6       &   9       & 0.04 &   170               \\
  $C$   & 1.8       &  6       &   9       & 0.01 &   170               \\
  $D$   & 1.8       &  6       &   9       & 0.02 &   200               \\ 
 \end{tabular} 
 \end{center} 
 \caption[Parameters used for simulations $A-D$]
 {Parameters used to construct simulated catalogues.
 The densities of 145, 170 and 200 sources per deg$^{2}$, correspond to 
 the surface densities of the present sample at flux density cutoffs of
 0.6, 0.5 and 0.4\,mJy respectively.}\label{tab_3} 
\normalsize  
\end{table}

\begin{equation}\label{eq_9a} 
 w(\theta)=A^{sim}_{w} \times \theta ^{-0.8}-\omega_{\Omega}, 
\end{equation} 

\noindent where the integral constraint, $\omega_{\Omega}$, is
0.54\,$A^{sim}_{w}$ for $\delta=-0.8$. The fitting is performed for angular
separations between 3 to 50\,arcmin. Additionally, for a given mock
catalogue, the angular correlation function is also calculated for nine
fields, distributed within the central region of the 50$\times$50 degree
area, each having the geometry of the Phoenix survey. The amplitude,
$A^{est}_{w}$, and the Poissonian uncertainty,  $\sigma_{Poisson}$,  are
calculated (see section \ref{sec_3})  for each of these smaller
fields and then compared to $A^{sim}_{w}$. The difference, $\Delta$,
between $A^{est}_{w}$ and $A^{sim}_{w}$ normalised to the Poisson standard
deviation is defined 

\begin{equation}\label{eq_10}
\Delta=\frac{|A^{sim}_{w}-A^{est}_{w}|}{\sigma_{Poisson}}.
\end{equation} 

\noindent For a given survival probability, $f$, and surface density of
objects (corresponding to a  flux density cutoff) six artificial catalogues
are generated. The cumulative distribution of $\Delta$ values (Figure
\ref{fig_7}), shows that  in $\approx70\%$ of the trials, $A^{est}_{w}$ lies
within   $\approx 2\sigma _{Poisson}$ from $A^{sim}_{w}$.  For a Gaussian 
distribution, this is the definition of one standard deviation. Therefore,
the simulations carried out here indicate that the angular correlation
amplitudes determined from a radio survey with similar characteristics
(i.e. geometry, surface density of objects) to those of the Phoenix survey,
are representative of the faint radio population amplitudes within two
Poisson standard deviations. This error is smaller than that estimated by
the bootstrap resampling technique ($\approx3\,\sigma_{Poisson}$; see
section \ref{sec_3}), suggesting that cosmic variance is not significantly
affecting the present study. 
However, it is likely that the simulations presented here do not correctly
model  the  higher order correlations produced by gravitational
instabilities, on which the $w(\theta)$ errors depend. Nevertheless, the 
Poisson errors estimated here are independent of higher order correlations
and are only meaningful when compared relative to the Poisson
uncertainties  of the real data-set.  
In the rest of this study we retain the bootstrap uncertainties as the
formal errors, since this method has been widely used to assess the
accuracy of $w(\theta)$.   

\section{The Model} 
Let us assume that the evolution of clustering of galaxies with redshift is 
parametrised by $\epsilon$ such that their spatial correlation function 
is (Phillipps {et al.} 1978) 
 
\begin{equation}\label{eq_11} 
\xi(r,z)=\biggl(\frac{r}{r_{o}}\biggr) ^{-\gamma}(1+z)^{-(3+\epsilon)}, 
\end{equation} 
 
\noindent where $r$ is the proper length and $r_{o}$ the correlation length for 
$z=0$, corresponding to the scale where the clustering becomes non-linear.  
The clustering evolution parameter has the property that if $\epsilon\geq
0$ the physical length of a typical cluster  reduces with time, resulting
in a growth of clustering  in proper coordinates. Values of $\epsilon<0$
result  in reduction in clustering strength  with time. The case of
$\epsilon\approx -1.2$ corresponds to clusters of fixed size in co-moving
coordinates.   
 
If the redshift distribution $dN/dz$ of galaxies is known, then the angular 
correlation function  $w(\theta)$ can be related to $\xi(r)$ by  
the Limber equation (Limber 1953 ; Phillipps {et al.} 1978) 

\begin{eqnarray}\label{eq_limber}
& w(\theta)=Cr_{o}^{1-\gamma} \theta^{-(\gamma-1)} 
\int_0^\infty D(z)^{1-\gamma}g(z)^{-1} & \nonumber \\ 
& \times (1+z)^{-(3+\epsilon)} 
\biggl(\frac{dN}{dz}\biggr) ^{2}\,dz  &  \nonumber \\ 
&\times \biggl[\int_0^\infty \frac{dN}{dz}\,dz \biggr] ^{-2}.  & 
\end{eqnarray} 
 
\noindent Here, $D(z)$, $g(z)$ and $C$ are defined as 

\begin{equation}\label{eq_13} 
D(z)=\frac{c}{H_{o}q_{o}^{2}} \frac{q_{o}z+(q_{o}-1)(\sqrt{1+2q_{o}z}-1)}{(1+z)^2}, 
\end{equation} 
 
\begin{equation}\label{eq_14} 
g(z)=\frac{c}{H_{o}} \frac{1}{(1+z)^{2} \sqrt{1+2q_{o}z}}, 
\end{equation} 
 
\begin{equation}\label{eq_15} 
C=\sqrt{\pi} \frac{\Gamma((\gamma-1)/2)}{\Gamma(\gamma /2)}. 
\end{equation} 
 
\noindent It is evident that the exponent $\gamma$ of the spatial
correlation function is related to the index $\delta$  of the angular
correlation function via the relation  $\gamma=\delta+1$.  
 
Unlike optically selected galaxies, the redshift information of the faint
radio population is sparse and  restricted only to sources with optical
counterparts that are relatively bright. Therefore, to predict the
amplitudes of the correlation function using equation (\ref{eq_limber}) 
for different input parameters ($\epsilon$, $r_{o}$, $\gamma$), we need to
use  a model radio luminosity function (RLF) to estimate their redshift
distribution. In the present work the evolving radio luminosity functions
of  Dunlop \& Peacock (1990) at 2.7\,GHz (Model 1) and  Rowan-Robinson {et
al.} (1993) at 1.4\,GHz (Model 2),  extrapolated  to faint flux densities, 
are used.   
 
\begin{figure} 
\centerline{\psfig{figure=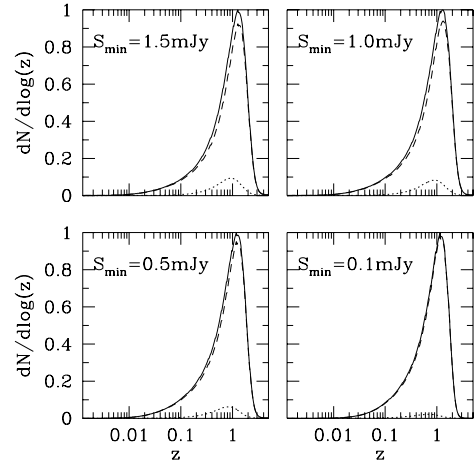,width=0.45\textwidth,angle=0}} 
 \caption {Redshift distributions for different flux density cutoffs using 
 the radio luminosity function introduced by Dunlop \& and Peacock
 (1990). Dashed line: Steep spectrum radio sources; Dotted line: Flat
 spectrum radio sources ; Solid line: both populations. The normalisation  
 of the redshift distribution is arbitrary.}\label{fig_8}
\end{figure}

Dunlop \& Peacock (1990) considered the free form luminosity  function of
both flat and steep spectrum radio galaxies. Their proposed evolutionary  
model implies that the radio sources of higher power evolve faster, with
the  evolution of flat and steep spectrum radio sources being similar.  
Here we employ their RLF1 model (using their RLF2, RLF3 and RLF4 models
give similar results) shifted to 1.4\,GHz and  converted to W\,Hz$^{-1}$.  
The transformation from 2.7\,GHz to 1.4\,GHz was carried out assuming a
power law spectral index $\alpha$ ($S_{\nu} \propto \nu ^{-\alpha}$) of
$0.8$ for steep spectrum and $0.0$ for flat spectrum radio sources.  
The redshift distributions at different flux limits for Dunlop \& Peacock  
models are shown in Figure \ref{fig_8}. The median of the redshift
distribution for flux limits of 0.1, 0.5 and 1\,mJy is 0.84, 0.85 and 0.91
respectively.  
 
\begin{figure} 
\centerline{\psfig{figure=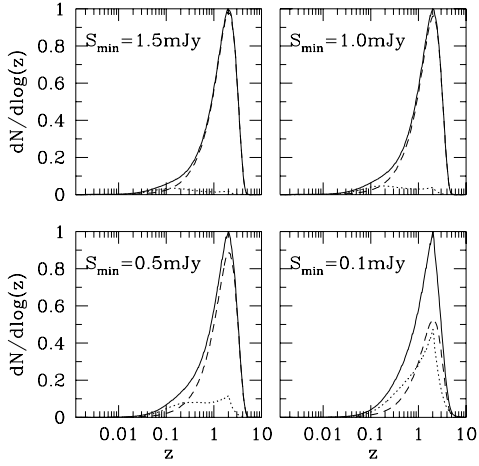,width=0.45\textwidth,angle=0}} 
 \caption{Redshift distributions for different flux density cutoffs using
 the radio luminosity function described by Rowan-Robinson {et al.}
 (1993). Dashed line: Ellipticals; Dotted line: Spirals; Solid line: both
 populations. The normalization of the redshift distribution is
 arbitrary.}\label{fig_9} 
\end{figure}

Rowan-Robinson {et al.} (1993) divided the radio population into
spirals, which they showed to be indistinguishable from the starburst
galaxies detected at 60\,$\mu$m by IRAS, and ellipticals. For the former
population we employ Saunders {et al.} (1990) `warm' IRAS galaxy
component luminosity function at 60\,$\mu$m translated to 1.4\,GHz using
the empirical relation derived by Helou {et al.} (1985).  For
ellipticals, following Rowan-Robinson {et al.} (1993), we use Dunlop \&
Peacock (1990) pure luminosity evolution  model RLF6 (parameters from their
Table C3). To match the source counts and the redshift distribution of the
optically brighter radio sources,  pure luminosity evolution for the spiral
galaxy component  is invoked as described in Rowan-Robinson {et al.}
(1993) and  Hopkins {et al.} (1998). The predicted redshift
distributions at  different flux density limits are plotted in Figure
\ref{fig_9}. The median of the  redshift distribution for flux limits of
0.1, 0.5 and 1\,mJy is 1.34, 1.37 and 1.44 respectively, significantly
larger than the median values predicted by the model developed  by Dunlop
\& Peacock (1990). Thus, the correlation amplitudes predicted by this RLF
model are lower compared to the previous one, for the same parameters
$r_{o}$, $\epsilon$.  
 
Using equation (\ref{eq_limber}) we  predict the amplitudes of $w(\theta)$
for different values of   $\epsilon$, $r_{o}$, $\gamma$ and for  different
radio luminosity functions. The value of $\gamma$ is fixed to either 1.8 or
2.1, with the  models calculated for each value of $\gamma$ independently.
Upper limits for the clustering length $r_{o}$ for $\epsilon=-1.2$, are
listed in Table \ref{tab_5}.  The error estimates of $r_{o}$ are
calculated from the 3$\sigma_{Poisson}$ correlation amplitude uncertainties
(Table \ref{tab_2}).

\begin{table} 
 \footnotesize 
 \begin{center} 
 \begin{tabular}{ccccc}  

 & \multicolumn{2}{c}{$r_{o}^{\gamma=(2.1)}$\,h$^{-1}$\,Mpc} &  
   \multicolumn{2}{c}{$r_{o}^{\gamma=(1.8)}$\,h$^{-1}$\,Mpc} \\ 
 $S_{1.4}$ (mJy) & model 1 & model 2 & model 1 & model 2 \\ 

 $>0.4$  &  $4^{+3}$   & $5^{+3}$    & $6^{+6}$ & $7^{+9}$ \\ 

 $>0.5$  &  $7^{+2}$   & $9^{+3}$    & $9^{+6}$ & $11^{+9}$  \\ 

 $>0.9$  &  $8^{+4}$   & $10^{+5}$   & $11^{+8}$ & $13^{+10}$\\

 $[0.4,0.9]$  &  $7^{+8}$  & $9^{+6}$ & $9^{+12}$ & $12^{+12}$\\ 
 
 \end{tabular} 
 \end{center} 
 \caption[Radio clustering length estimates]
 {Upper limits of the clustering length, for $\epsilon$=-1.2 and for
 two RLF models. Model 1: Dunlop \& Peacock (1990) RLF; model 2: RLF
 described by Rowan-Robinson {et al.} (1993). The superscripts are the
 upper limits to the errors in the $r_{o}$ estimates, corresponding to
 the bootstrap uncertainties of the angular correlation
 amplitude.}\label{tab_5} 
\end{table}

\begin{figure}
\centerline{\psfig{figure=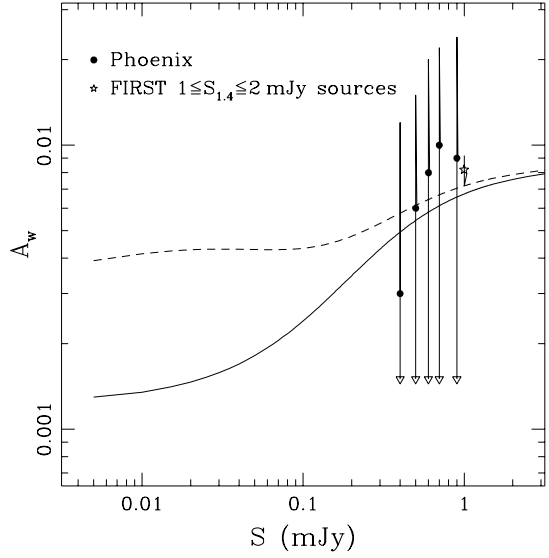,width=0.45\textwidth,angle=0}} 
 \caption[Relation between $A_{w}$ and $S_{1.4}$ for models $A$ and $B$]
 {Predicted amplitude of the angular correlation function at
 different flux density cutoffs, assuming that the radio population consists
 of starbursts and elliptical galaxies with different clustering properties
 Solid line: $r_{o}=5$\,h$^{-1}$\,Mpc for starbursts,
 $r_{o}=11$\,h$^{-1}$\,Mpc for ellipticals (Model $A$); Dashed line: 
 $r_{o}$=11\,h$^{-1}$\,Mpc for both populations (Model $B$). The clustering
 evolution parameter is taken to be $\epsilon$=-1.2 and $\gamma$=1.8 for
 both populations. Also shown are the estimated amplitudes from this work
 (filled circles; upper limits) and for sources with $1\le
 S_{1.4}\le2$\,mJy detected in the FIRST survey.}\label{fig_10} 
\end{figure}

\section{Discussion}\label{sec_6}

Peacock \& Nicholson (1991) have shown that apparently bright radio
sources ($S>0.5$\,Jy) in the redshift range $0.01<z<0.1$ are spatially
correlated, following a power-law behaviour $\xi=(r/11$h$^{-1}$
Mpc$)^{-1.8}$.  Loan, Wall \& Lahav (1997), using two large-area 4.85\,GHz
radio surveys covering $70\%$ of the sky, conclude that the 2-D
distribution of  radio sources brighter than 50\,mJy is consistent with a
correlation length  in the range $13<r_{o}<18$ \,h$^{-1}$\,Mpc and an evolution 
parameter $-1.2<\epsilon <0$. This result is based on the RLF models
developed by Dunlop \& Peacock (1990) and a value of $\gamma$=1.8. In a
separate study, Cress {et al.}  (1996) estimated the angular
correlation function of the FIRST  radio survey for sources with flux
density $S_{1.4}>$1\,mJy. Assuming $\epsilon=-1.0$ and $\gamma=2.1$, they found
$r_{o}$=6-8\,h$^{-1}$\,Mpc  (Cress {et al.} 1997).  

The $r_{o}$ upper limits estimated in this study at different flux density
cutoffs agree with the estimates derived from brighter radio samples (Cress
{et al.} 1997; Loan, Wall \& Lahav 1997). Because of the small number
of sources in each of the independent flux-limited sub-samples
($\approx$318) the uncertainties are large  and thus we cannot
conclude from the present  sample  if there is a 
change in the clustering properties of radio sources with flux density.
Figure \ref{fig_10} compares the correlation amplitude upper 
limits from this study with those calculated for the radio sources
detected in the FIRST radio survey 
with $1\le S_{1.4}\le2$\,mJy (Cress {et al.} 1996). The agreement is good, 
even within the Poisson errors.  

In recent years, deep radio surveys, (e.g. Windhorst {et al.} 1985;
Fomalont {et al.} 1997; Hopkins  {et al.} 1998) have shown a
flattening in the slope of the normalised number counts at sub-mJy levels,
revealing  an excess of faint radio sources over the `normal' radio
population of giant ellipticals and QSOs. To address this problem, models
invoking  strong evolution of either spiral galaxies (Condon 1989) or  
star-forming IRAS galaxies (Rowan-Robinson {et al.} 1993; Hopkins {et al.}
1998) have been combined with RLF models of the `normal' radio    
population. This scenario is supported by photometric and spectroscopic
studies revealing that the sub-mJy  radio sources can be identified with
galaxies exhibiting evidence of increased star formation (Benn {et al.}
1993; Windhorst {et al.} 1985). Furthermore, studies of local galaxies
with enhanced  star formation (late spirals, IRAS galaxies, HII galaxies),
has shown  that those objects are more weakly clustered
($r_{o}\approx$2-4\,h$^{-1}$\,Mpc, for $\gamma\approx1.5-1.7$; Davis \&
Geller 1976; Giovanelli et {al.} 1986; Saunders, Rowan-Robinson \&
Lawrence 1992 ) than E/S0  galaxies. This implies that the sub-mJy
population could be more weakly clustered than the E/S0 objects that host
the majority of apparently brighter radio sources. Support for this view
was advanced by Cress  {et al.} (1996) who found that the slope of  the
angular correlation function of the sources  with 1$<S_{1.4}<2$\,mJy is
flatter compared to that found  for $S_{1.4}>$3\,mJy. They interpret this
result as being due to the increased contribution from starburst galaxies
at lower flux density limits, which have flatter angular correlation
functions compared to ellipticals, which dominate at brighter flux
densities. A similar argument is used by Peacock (1997) to explain the
apparent conflict between the value of $r_{o}=$6.5\,h$^{-1}$\,Mpc found
for a sample of radio  sources brighter than 2.5\,mJy and the value 
$r_{o}=$11\,h$^{-1}$\,Mpc found by Peacock \& Nicholson (1991) for sources
brighter than 500\,mJy and redshifts in the range $0.01<z<0.1$. 

To explore further the implications of the two scales of clustering, and in
particular to explore the potential to eliminate competing models by
observations, the amplitude of the angular correlation function was
estimated, by adopting a simple model (Model $A$) in which the radio
population  consists of  two components, one dominating at brighter
($>$1\,mJy) fluxes, with correlation length $r_{o}=11$\,h$^{-1}$\,Mpc
($\gamma$=1.8) and the other dominating at sub-mJy levels with
$r_{o}=5$\,h$^{-1}$\,Mpc ($\gamma$=1.8) similar to that found for local
starburst galaxies. Any cross-correlation between the two radio
populations is ignored. The clustering evolution parameter is taken to be
$\epsilon$=-1.2 and our RLF model 2 is employed to predict  the redshift
distribution of the two radio populations at faint flux densities. In
Figure \ref{fig_10} we plot the flux density cutoff against the amplitude
of the angular correlation function calculated from model $A$. For
comparison, the expected relation assuming the same value of
$r_{o}=11$\,h$^{-1}$\,Mpc ($\gamma$=1.8) for the two populations  (Model
$B$), is also plotted, along with the upper limits for the  angular
correlation amplitudes, $A_{w}$, calculated  in the present study for
different flux density cutoffs. The uncertainties are too large to allow
discrimination  between the two models. 

\begin{figure} 
\centerline{\psfig{figure=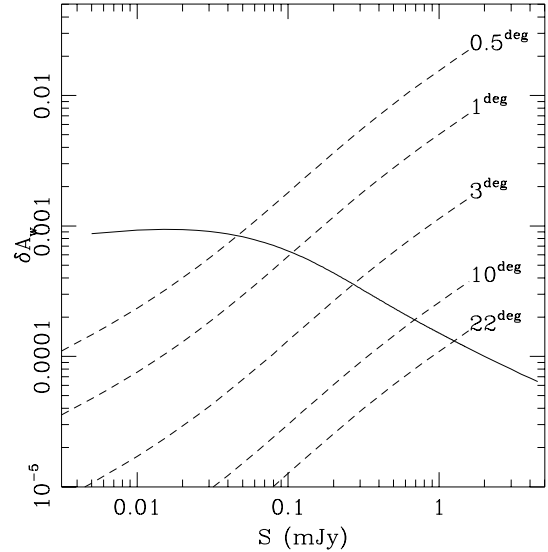,width=0.45\textwidth,angle=0}} 
\caption {Flux density cutoff versus the estimated uncertainty in the
 amplitude of the angular correlation function. Dashed lines correspond to
 surveys of circular geometry with different radii. The Phoenix survey has
 a radius of 1$^{\circ}$, while the FIRST radio survey is subtending, at the
 moment, an area corresponding to a radius of $\approx$22$^{\circ}$. The
 solid  line represents the locus of 3$\sigma$ confidence level 
 discrimination between models $A$ and $B$ described in the
 text.}\label{fig_11}   
\end{figure}

It is an interesting exercise to predict the depth and the solid angle
subtended  by a radio survey that would reveal  at a 3$\sigma$ significance
level if the clustering properties of faint radio sources were
significantly different  from those of the brighter ones as a result of the
changing population. This is accomplished by estimating the uncertainty in
$A_{w}$ for a survey of a 
given solid angle and   completeness limit, as described in 
Appendix A. This then is  compared to the difference $\Delta A_{w}$
between the   correlation amplitudes predicted at the same flux density
cutoff from Models $A$ and $B$. To discriminate between the two models at a
3$\sigma$ confidence level, the uncertainty  in  $A_{w}$ should be 3 times
smaller than $\Delta A_{w}$.  The results are shown in Figure
\ref{fig_11}, where the flux density cutoff is plotted against the
uncertainty in $A_{w}$ for different survey areas.  The solid line delimits
the area in the parameter space in which $\delta A_{w}$ is at least 3 times
smaller than $\Delta A_{w}$ and hence defines the locus of 3$\sigma$
confidence level discrimination between the  simplified models $A$ and
$B$. We conclude that either a very  deep survey ($S_{lim}$=40\,$\mu$Jy)
over $\approx1$\,deg$^{2}$ or a survey over a larger area at a brighter
limit (e.g. at $S_{lim}$=0.2\,mJy over $\approx27$\,deg$^{2}$) is required to
discriminate between models $A$ and $B$ and thus reveal if there exists a
weakly clustered radio population  at faint flux densities. Our analysis
shows that the FIRST radio survey, ($S_{lim}=1$\,mJy) covering, at  the
moment, an area  of $\approx1500$\,deg$^{2}$, is also appropriate for this
purpose. 
 
\section{Conclusions} 
Using a deep, homogeneous survey covering an area of $\sim3$\,deg$^{2}$, we
have investigated the  clustering properties of faint (sub-mJy) radio
sources by calculating the angular correlation function of the sample.  
Simulations have been carried out to investigate the significance of our
results. It has been demonstrated that the amplitudes estimated here imply
a non-uniform distribution of radio sources, albeit at the 2$\sigma$
significance level. 

Furthermore, we find that the correlation amplitude upper limits
estimated here, despite the large uncertainties,  are consistent
with the results derived in other studies from larger samples at brighter
flux density limits. Spectroscopic and photometric studies of sub-mJy radio
sources show that the majority have properties similar to starburst 
galaxies. Since local galaxies, with increased star formation, exhibit
weaker clustering than average, it is tempting to link the clustering
properties of the sub-mJy population with those of local starburst
galaxies. However, the present data, due to small number statistics, cannot
reveal a change in radio source clustering properties with flux
density. Nevertheless, we have demonstrated that future surveys having
sensitivities at $\mu$Jy flux density levels and covering large areas on
the sky, will be suitable to explore possible changes of the correlation
amplitude with flux density.

\begin{acknowledgements} 
AG is supported by a scholarship from  the State Scholarships Foundation of
Greece (S.S.F.). The work of LC and AH is supported by the Australian
Research Council and the Science Foundation for Physics within the
University of Sydney. We thank the referee for useful comments and
suggestions that improved this paper.
\end{acknowledgements}

\appendix 

\section{Angular Correlation Function Amplitude Uncertainty}\label{app}

The uncertainty in the amplitude of the angular correlation function of 
a fiducial radio survey  with limiting flux density $S_{lim}$, subtended 
over a solid angle $\Omega$, is estimated here. To simplify the procedure, 
the following assumptions are made:  

\begin{enumerate}

 \item the area subtended by the survey has a circular geometry;

 \item the angular correlation function of radio sources is a power law of
 the form $w(\theta)=A_{w}\theta^{-\delta}$, with the index $\delta$ being
 fixed; 

 \item the values of $w(\theta)$ at different angular separations are
 uncorrelated; 

 \item The uncertainty in $w(\theta)$ follows Poisson statistics  (Peebles
 1980)  and is given by 

 \begin{eqnarray} 
 &\delta w(\theta) & =\sqrt{\frac {1+w(\theta)}{DD}},  \nonumber \\ 
 &                 & =\frac {1}{\sqrt{\frac{1}{2}N\,(N-1) 
                   <d\Omega>/\Omega}},  
 \end{eqnarray} 

 \noindent where DD is the number of data-data pairs with angular separations
 $\theta$ and $\theta+d\theta$,  $N$ is the number of sources above the
 flux density limit and $<d\Omega>$ is the mean value of solid angle
 subtended by annuli with inner and outer radii $\theta$ and
 $\theta+d\theta$ respectively. The value of $<d\Omega>$ is affected by the
 field boundaries. In the  present study, this is estimated by creating
 simulated catalogues of sources distributed randomly over the area of the
 survey and calculating the expected number of random points, RR, with
 separations $\theta$ to $\theta+d\theta$. This is given by

 \begin{equation} 
 RR =\frac{1}{2}N_{R}(N_{R}-1)\frac{<d\Omega>}{\Omega},   
 \end{equation} 

 \ noindent where $N_{R}$ is the number of sources in the simulated
 dataset.The assumption of Poisson errors underestimates the uncertainty in
 $A_{w}$ (Bernstein 1994). 
\end{enumerate}

The assumptions (iii) and (iv) allow us to use least squares fitting
techniques to estimate  the uncertainty in $A_{w}$ 

\begin{equation}\label{eq_D3}
 \delta A_{w}=\biggl (\sum_{i=\theta_{min}}^{\theta_{max}} \biggl(\frac{\theta_{i}^{-\delta}-\omega_{\Omega}}{\delta w(\theta_{i})}\biggr)^{2} \biggr)^{-\frac{1}{2}},  
\end{equation} 

\noindent where $\omega_{\Omega}$ is the integral constraint defined in
section \ref{sec_3}. The summation is carried out between a minimum
($\theta_{min}$)  and a maximum ($\theta_{max}$) angular separation within
equally  separated logarithmic bins  of width $\Delta$log$\theta=0.2$
($\theta$ in degrees). The former is taken to be 3$^{\prime}$  and the
latter is set  three times smaller than the radius of the survey, while it
cannot be larger than 1$^{\circ}$. This is because in real surveys one
needs to minimise the effects of field boundaries when estimating
$A_{w}$. Additionally, one is interested in the value of $A_{w}$ on small
scales, while for larger scales the angular correlation function may
deviate from a power law. For example in the optical wavelengths Maddox
{et al.} (1990a) found a break in $w(\theta)$ at scales $\theta\approx1.5''$.


\begin{thebibliography}{} 

\bibitem[1995]{becker} Becker R. H., White R. L., Helfand D. J.,  1995,
ApJ, 450, 559 
 
\bibitem[1995]{benn95} 
Benn~C.~R., Wall J. V., 1995, MNRAS, 272, 678

\bibitem[1993]{benn93} 
Benn~C.~R., Rowan-Robinson~M., McMahon~R.~G., Broadhurst~T.~J.,
\&~Lawrence~A., 1993,~MNRAS,~263,~98  

\bibitem[1995]{bern} Bernstein G. M., 1994, ApJ, 424, 569

\bibitem[1997]{carlberg} Carlberg R. G., Cowie L. L., Songaila A., Hu E.,
M., 1997, ApJ, 484, 538

\bibitem[1984]{condon84}  
Condon~J., 1984,~ApJ,~287,~461 
 
\bibitem[1989]{condon89} 
Condon~J.~J., 1989,~ApJ,~338,~13 
 
\bibitem[1996]{cress96} 
Cress~C.~M., Helfand~D.~J., Becker~R.~H., Gregg~M.~D., 
\&~White~L.~R., 1996,~ApJ,~473,~7 
 
\bibitem[1997]{cress97} 
Cress~C.~M. \& FIRST collaboration, 1997, astro-ph/9711232
 
\bibitem[1976]{davis76} 
Davis~M., \& Geller~M.~J., 1976,~ApJ,~208,~13 
 
\bibitem[1983]{davis83} 
Davis~M., Peebles~P.~J.~E., 1983,~ApJ,~267,~465 
 
\bibitem[1990]{dunlop}  
Dunlop~J.~S., Peacock~J.~A., 1990,~MNRAS,~247,~19 

\bibitem[1999]{george}  
Georgakakis A., Mobasher B., Cram L., Hopkins A., Lidman C., Rowan-Robinson
M., 1999, MNRAS, 306,708
 
\bibitem[1986]{giovanelli} 
Giovanelli~R., Haynes~M.~P., \&~Chincarini~G.~L., 1986, ApJ, 300, 77  

\bibitem[1994]{fisher} 
Fisher B. K., Davis M., Strauss A. M., Yahil A., Huchra J.,
MNRAS, 1994, 266, 50

\bibitem[1997]{fomalont} 
Fomalont E. B., Kellermann K. I., Richards E. A., Windhorst R. A.,
Partridge R. B., 1997, ApJ, 475, L5 
 
\bibitem[1993]{hamilton} 
Hamilton~A.~J.~S., ApJ~1993,~417,~19 
 
\bibitem[1993]{helou} 
Helou G., Soifer B. T., Rowan-Robinson M., 1985, ApJ, 298, L7 

\bibitem[1982]{hewett} 
Hewett P. C., 1982, MNRAS, 201, 867
 
\bibitem[1998]{hokins} Hopkins A., Mobasher~B., Cram~L., Rowan-Robinson~M., 1998,
MNRAS, 296, 839
  
\bibitem[1997]{hudon} 
Hudon J. D., Lilly J. S., 1996, ApJ, 469, 519  
 
\bibitem[1994]{infante94} 
Infante~L., 1994,~A\&A,~282,~353 
 
\bibitem[1995]{infante95} 
Infante~L., Pritchet~C.~J., 1995,~ApJ,~439,~565 
 
\bibitem[1995]{kooiman}  
Kooiman~B.~L., Burns~J.~O., Klypin~A.~A., 1995,~ApJ,~448,~500 
 
\bibitem[1993]{landy} 
Landy~S.~D., Szalay~S.~A., 1993,~ApJ,~412,~64 
 
\bibitem[1953]{limber} 
Limber~D.~N., 1953,~ApJ,~117,~134 

\bibitem[1986]{ling} 
Ling~E.~N., Frenk~C.~S., \& Barrow~J.~D., 1986,~MNRAS,~223,~21p  
 
\bibitem[1997]{loan} 
Loan~A.~J., Wall~J.~V, Lahav~O., 1997,~MNRAS,~286,~994 

\bibitem[1995]{loveday} 
Loveday J., Maddox S. J., Efstathiou G., Peterson B. A., 1995, ApJ, 442,
457 
 
\bibitem[1990a]{maddox90a} 
Maddox~S.~J., Efstathiou~G, Sutherland~W.~J.,   
\& Loveday~J, 1990a,~MNRAS,~242,~43 

\bibitem[1990b]{maddox90b} Maddox~S.~J., Efstathiou~G, Sutherland~W.~J., \&
Loveday~J, 1990b,~MNRAS,~242,~43

\bibitem[1998]{maglio} 
Magliocchetti M., Maddox S. J., Lahav O., Wall J. V., 1998, MNRAS, 300, 257
 
\bibitem[1979]{masson} 
Masson~C., 1979,~MNRAS,~188,~261 

\bibitem[1992]{mo} 
Mo H. J., Jing Y. P., Borner G., 1992, ApJ, 392, 452 

\bibitem[1985]{peacock85}  
Peacock~J.~A., 1985,~MNRAS,~217,~601 
 
\bibitem[1991]{peacock91} 
Peacock~J.~A., Nicholson~D., 1991,~MNRAS,~253,~307 

\bibitem[1997]{peacock97} 
Peacock~J.~A., 1997, proceedings of the KNAW colloquium, {\it  The most
distant radio galaxies}, Amsterdam, October 1997 astro-ph/9712068

\bibitem[1980]{peebles} 
Peebles P. J., 1980, The Large Scale Structure of the Universe (Princeton:
Princeton University Press) 
 
\bibitem[1978]{phillipps}  
Phillipps S., Fong R., Fall R. S., Ellis S. M., Macgillivray H. T.,
1978,~MNRAS,~182,~673  

\bibitem[1987]{oort87a} Oort M. J. A., 1987a, Ph.D. thesis, University of Leiden

\bibitem[1987]{oort87b} Oort M. J. A., 1987b, A\&AS, 71, 221

\bibitem[1999]{richards1999} Richards E. A., 1999, astro-ph/9908313

\bibitem{e17} Roche N., Eales S., Hippelein H., 1998a, MNRAS, 295, 946

\bibitem{e21} Roche N., Eales S., 1998b, astro-ph/9803331
 
\bibitem[\protect\citename{Rowan-Robinson\ }1993]{24} 
Rowan-Robinson~M., Benn~C.~R., Lawrence~A., McMahon~R.~G., \&
Broadhurst~T.~J., 1993,~MNRAS,~263,~123  
 
\bibitem[\protect\citename{Saunders\ }1992]{24a}  
Saunders~W., Rowan-Robinson~M., \& Lawrence~A., Efstathiou~G., Kaiser N.,
Ellis R. S., Frenk C. S. 1990, MNRAS,~242,~318  
 
\bibitem[\protect\citename{Saunders\ }1992]{25}  
Saunders~W., Rowan-Robinson~M., \& Lawrence~A., 1992, MNRAS,~258,~134 

\bibitem[\protect\citename{Shavier\ }1992]{25a}  
Shaver P. A., \& Pierre M., 1989 A\&A, 220, 35

\bibitem{c28} Soneira R. M., Peebles P. J. E., 1978, AJ, 83, 845

\bibitem{c4} Villumsen J. V., Freudling W., Da Costa L. N., 1997, ApJ,
481, 578 

\bibitem[\protect\citename{Webster\ }1976]{26} 
Webster~B.~L., 1975,~MNRAS,~175,~61 
 
\bibitem[\protect\citename{Widhorst\ }1985]{27} 
Windhorst~R.~A., Miley~K.~G., Owen~F.~N., Kron~R.~G., David~C.~K.,
1985,~ApJ~289,~494  

\bibitem[\protect\citename{Widhorst\ }1985]{27a} 
Windhorst~R.~A., Mathis D., Neuschaefer L., 1990, ASP Conf. Series, 10, 389

\bibitem[\protect\citename{Widhorst\ }1985]{27b} 
Windhorst R. A., Fomalont E. B., Partridge R. B., Lowenthal J. D., 1993,
ApJ, 405, 498

\end{thebibliography}
\end{document}